\documentclass{aa}  

\usepackage{graphicx}
\usepackage{txfonts}
\usepackage{color}

\begin{document}

   \title{Early-type dwarf galaxies with multicomponent stellar structure: Are they remnants of disc galaxies strongly transformed by their environment?}

   \author{J. Alfonso L. Aguerri
          \inst{1,2}
          }
   \institute{Instituto de Astrof\'{\i}sica de Canarias. E-38200, La Laguna, Tenerife, Spain
        \and
    Universidad de La Laguna, Dept. Astrof\'{\i}sica, E-38206 La Laguna, Tenerife, Spain \\
              \email{jalfonso@iac.es}
             }

   \date{Received ; accepted}

  \abstract
   {The surface brightness distribution of $\sim$30--40$\%$ of the early-type dwarf galaxies with $-18 \leq M_{B} \leq -15$ in the Virgo and the Coma clusters is fitted by models that include two structural components (S\`ersic + exponential) as for bright disc galaxies.}
   {The goal of the present study is to determine whether early-type dwarf galaxies with a two-component stellar structure in the Virgo and the Coma clusters are low-luminosity copies of bright  disc galaxies  or are the remnants of  bright galaxies strongly transformed by cluster environmental effects.}
   {I analysed the location of bright disc galaxies and early-type dwarfs in the $r_{b,e}/h$-- $n$ plane. The location in this plane of the two-component dwarf galaxies was compared with the remnants of tidally disrupted disc galaxies reported by numerical simulations.}
   {Bright unbarred disc galaxies show a strong correlation in the $r_{b,e}/h$--$n$ plane. Galaxies with larger S\`ersic shape parameters show a higher $r_{b,e}/h$ ratio. In contrast, two-component early-type dwarf galaxies do not follow the same correlation. A fraction ($\sim$55\%) of them are located outside the locus defined in this plane by having 95$\%$ of bright disc galaxies. This distribution indicates that they are not a low-mass replica of bright disc galaxies. The different location in the $r_{b,e}/h$-- $n$ plane of two-component early-type dwarfs and bright galaxies can be qualitatively explain whether the former are remnants of disc galaxies strongly transformed by tidal processes.
   }
   {The progenitors of $\sim$20-25\% of early-type dwarf galaxies with $-18 \leq M_{B} \leq -15$ in the Virgo and Coma clusters could be  bright disc galaxies transformed by effects of the environment. These tidally transformed galaxies can be selected according to their location in the $r_{b,e}/h$--$n$ plane. }


\titlerunning{Bright disc galaxies transformed by enviromental effects}

   \maketitle
%

\section{Introduction}

Automatic galaxy classifications use different physical parameters to classify galaxies in several morphological types. We can mention the galaxy colours, bulge-to-disc ratio, luminosity, light concentration, and asymmetry as some of the parameters that have been extensively used in the literature. Galaxies can be divided into  two large groups according to their luminosity: bright ($M_{B}<-18.0$) and dwarf $(M_{B}>-18.0)$ galaxies. Historically, the cut at $M_{B}=-18.0$ has been caused by the change in the behaviour of certain scaling relations of galaxies at this luminosity \citep[see][]{binggeli84a,kormendy1985}. New data have questioned the reality of this discontinuity \citep[see e.g.][]{jerjen1997,graham2003,gavazzi2005,aguerri2005a,cote2006}. Nevertheless, it is well documented that both bright and dwarf galaxies  contain systems showing different structures, colours, stellar contents, kinematics, and probably origins. 

Dwarf galaxies are the most abundant type of galaxies in the Universe \citep[see][]{phillips1998}. The study of their properties is of great interest because, according to the cold dark matter (CDM) theory, the structure formation in the Universe was built up by merging small sub-units into larger ones. Dwarf galaxies are therefore the building blocks of large and massive galaxies. In addition, their small gravitational potential wells render them fragile against tidal interactions. This implies that they are ideal objects for studying environmental processes and mass assembly in galaxy clusters.

According to their stellar population content, dwarf galaxies can be classified as star-forming \citep[dwarf irregulars and blue compact dwarfs; e.g.][]{thuan81,vanzee00,amorin09} and quiescent \citep[dwarf ellipticals, compact ellipticals, and dwarf spheroidals; e.g.][]{binggeli84b,kormendy09}. The former are more abundant in the field, and the latter are located in high-density environments as galaxy groups or clusters  \citep[see e.g.,][]{hogg2004,sanchezjanssen2008}. This dichotomy can also be observed in the galaxy luminosity function. Thus, the faint end of the luminosity function of field galaxies is formed mainly by blue and low-mass galaxies \citep[e.g.\ ][]{blanton2005,monterodorta2009}. In contrast, red dwarf galaxies dominate in number the faint end of the luminosity function of galaxies in nearby clusters \citep[e.g.\ ][]{depropris2003,popesso2006,agulli2014}. 

In this study I denote the quiescent dwarf systems located mainly in galaxy clusters as early-type dwarf galaxies. I will also denote those systems showing distinct bulge-to-disc ratios as bright disc
galaxies. This includes S0, early, and late-type spiral galaxies.

That early-type dwarfs are located in high-galaxy density environments has been interpreted as a link between their origin and the environmental processes that galaxies undergo in clusters.  Several physical processes taking place in clusters can produce strong galaxy evolution \citep[see][]{boselli2006}  although some affect only the gas content of the galaxies. We may mention ram-pressure stripping \citep[see][]{gunn1972,quilis2000,bekki2009}, starvation \citep[see][]{larson1980,bekki2002}, and strangulation \citep[see][]{kawata2008}. Other processes affect both their gas and stellar contents; these include harassment \citep{moore1996}, tidal interactions \citep{merritt1984,aguerri2009}, and mergers. These processes can strongly change the morphology \citep[see][]{aguerri2004, aguerri2005a, lisker2006a, mendezabreu2010}, kinematics \citep[see][]{pedraz2002, geha2003, toloba2009}, and stellar content \citep[see][]{haines2006, smith2009} of galaxies on time scales smaller than a few Gyr. All of these physical mechanisms ensure that part of the stellar and/or gas content of the galaxies are finally located as free-floating objects within the cluster potential. The galactic material removed from galaxies constitutes the so-called intracluster light observed in some nearby galaxy clusters  \citep[e.g.\ ][]{arnaboldi02, aguerri05b, gerhard05, castrorodriguez09}. It is still a matter of debate which of these physical mechanisms drives the evolution of dwarf galaxies in high-density environments \citep[see][for a recent review]{boselli2014}.

The quenching of star formation in galaxies located in high-density environments has been reported well in the literature \citep[e.g.\ ][]{lewis02}.  Some observational studies suggest that early-type dwarfs can result from the recent migration of star-forming faint galaxies through the green valley. This migration would be due to the abrupt and fast truncation of their star formation rate produced by the sweeping out of their cold gas by the ram-pressure effect.  Within this framework, the progenitors of early-type dwarf galaxies would be dwarf systems with active star formation that was quenched by the sweeping out of their cold gas content by the interaction of the galaxy with the hot intracluster medium \citep[see][]{boselli2008,gavazzi2010}. The lines of observational evidence for this argument include (i) the similar faint-end slope of the galaxy luminosity function observed in nearby galaxy clusters and the field \citep[see e.g.,][]{boselli2011, agulli2014}; (ii) the relation of the gas content, colour, and environment found in clusters \citep[e.g.\ ][]{sanchezjanssen2008, gavazzi2013}; (iii) the age and metallicity spread observed in dwarf galaxies of nearby clusters, together with the correlation between age, metallicity, and clustercentric distance \citep[see][]{boselli2008, smith2009, toloba2009, koleva2013}; the globular cluster content \citep[see][]{sanchezjanssen12}; and the different kinematics properties between galaxies located in the inner and the outermost regions of the clusters \citep[see][]{toloba2009}.

The family of early-type dwarf galaxies is so complex that it is unlikely that their formation and evolution can be explained by a single process. A fraction of  early-type dwarf galaxies in clusters show complex structure, rotation, or the presence of gas, dust, and star formation in their centres \citep{lisker2006a,lisker2006b,lisker2007}.  These properties indicate that these objects are low-mass disc galaxies. The origin of these early-type dwarfs  is still unclear: Are they low-mass copies of bright disc galaxies with quenched star formation by ram-pressure stripping? Or are they the remnants of bright disc galaxies strongly transformed by environmental processes? In the first case, the  stellar scaling relations and kinematic properties of low-mass disc galaxies would be similar to those of bright disc galaxies because the ram-pressure stripping only affects the gas content of galaxies, leaving the stellar component almost unchanged \citep[see][]{kenney2004,kronberger2008}. In the other case, it is expected that low-mass discs have different scaling relations and kinematic properties from bright disc galaxies. In this case, tidal interactions strongly perturb their stellar distribution and/or stellar kinematics  \citep{aguerri2009}. 

The $N$-body simulations show that tidal interactions can  undergo major structural transformations on galaxies in clusters \citep[see][]{mastropietro2005, aguerri2009, benson2014}. Several fast encounters of bright/massive bulge-disc galaxies with the cluster potential or the cluster substructure can remove up to 80--90$\%$ of their total mass \citep[see][]{penarrubia2008,aguerri2009}.  This mass truncation also produces significant changes in their structural parameters, such as effective radius ($r_{b,e}$) and S\`ersic shape ($n$) parameters of the bulges and the scale length ($h$) of the discs, although the bulge-disc structure is kept in the remnants \citep[see][]{aguerri2009}.


Studies of structural parameters in early-type dwarf galaxies show that, for most of them, their  surface brightness distributions can be properly fitted by a single S\`ersic profile with shape parameters $n \approx 1$ \citep[see e.g.][]{binggeli91,barazza03}. Nevertheless, a fraction of early-type dwarf galaxies need more than one component to fit their surface brightness distributions \citep[see][]{aguerri2005a,janz2014}.  These early-type dwarf galaxies with two structural components are usually called dS0. In contrast, we call dwarf ellipticals (dE) those early-type dwarfs with only one structural component. In a study of 99 dwarf galaxies in the Coma cluster, Aguerri et al. (2005) found that about 30$\%$ of them have two structural components. More recently, \cite{janz2014} have shown that two-thirds of their dwarf galaxies in the Virgo cluster show a multicomponent structure including bulges, discs, bars, and lenses. 

The aim of the present paper is to analyse the structural parameters of a sample of dS0 galaxies located in the Virgo and Coma clusters within the framework of the simulations
of Aguerri \& Gonzalez-Garcia (2009). In particular, we study the location of the dS0 galaxies in the $r_{b,e}/h$--$n$ plane in comparison with that of the bright disc galaxies and the remnants of tidally disrupted massive disc galaxies as revealed by numerical simulations. According to \cite{aguerri2009}, it is in this plane that the remnants of strongly harassed disc galaxies are located in a different region from bright disc galaxies. 

The structure of the paper is as follows. Section 2 contains the description of the sample used in this work. The results are shown in Section 3. The discussion and conclusions are given in Sections 4 and 5, respectively. The cosmological parameters used in the present work are $H_{0} = 75$ km s$^{-1}$ Mpc$^{-1}$, $\Omega_{m} = 0.3$, and $\Omega_{\Lambda} = 0.7$.

\section{The galaxy samples and the galaxy structural parameters}

The sample of galaxies studied in this paper was divided into two subsets: bright and dwarf galaxys.  The structural parameters of the galaxies in both samples were obtained from several studies in the literature. 

\subsection{The bright galaxy sample}

The structural parameters of the bright disc galaxies were obtained from: 

\begin{itemize}
\item The sample from M\'endez-Abreu et al. (2008).  This sample of galaxies is formed by 148 galaxies with intermediate inclination ($i < 65^\circ$) and classified as discs with Hubble types from S0 to Sb. The structural parameters of the galaxies were obtained from 2MASS images in the J band
\item The sample of Graham (2003). This sample consists of 86 spiral galaxies from the local and volume-limited sample of de Jong \& van der Kruit (1994). The structural parameters were fitted on photometrical images observed through $B$, $R$, $I$, and $K$ bands.
\item Galaxies from Mollenhoff \& Heidt (2001), who obtained the structural parameters for 40 bright spiral galaxies with Hubble types from Sa to Sc. The best-fitting $J$, $H$, and $K$ structural parameters are given in this work.
\item Galaxies from MacArthur et al. (2003).  This sample is formed by 121 galaxies with Hubble types ranging from Sab to Sd and  intermediate inclination ($i<60^\circ$). Structural parameters of the galaxies in $B$, $V$, $R$, and $H$ bandpasses are provided.
\end{itemize}

A common feature of these samples is that the structural parameters of the galaxies were obtained by fitting only two components (bulge and disc) to the surface brightness of the galaxies. All these studies used the same mathematical expressions for models describing the two galaxy components. In particular, the surface brightness profiles of the bulge was modelled by the S\`ersic's law, given by

\begin{equation}
I(r)=I_{b,e} \times 10^{-b_{n}[(r/r_{b,e})^{1/n}-1]}
\end{equation}

\noindent where $r_{b,e}$, $I_{b,e}$, and $n$ are the effective radius, surface brightness at $r_{b,e}$, and  the so-called S\`ersic shape parameter, respectively. The value of $b_{n}$ is coupled to $n$ so that half of the total flux of the profile is enclosed within $r_{b,e}$. It can be approximated by $b_{n} = 0.868n - 0.142$ \citep[see][]{caon93}. This mathematical law has been extensively used in the literature to fit the surface brightness profiles of bulges across the Hubble sequence \citep[e.g.\ ][]{andredakis1995, trujillo2002, aguerri2004, simard2011, meert2015}.

 The discs of the galaxies were fitted using an exponential law \citep[][]{freeman1970} given by

\begin{equation}
I(r)=I_{d,0} \times e^{-r/h}
,\end{equation}

\noindent $I_{d,0}$ and $h$ being the central disc surface brightness and the disc scale length, respectively.

Only galaxies with $M_{B}<-18.0$ were considered in the bright galaxy sample.\footnote{I used only those galaxies with measured $B$-band magnitudes. These magnitudes were obtained from the articles quoted in this paper or from the Nasa Extragalactic Database (NED)}. To minimize the differences in the structural parameters due to fits in different wavelengths, only those galaxies with structural parameters obtained in $J$, $H$, or $K$ bands were considered. This means that the total number of bright galaxies taken into account in this study  is 297.

\subsection{The dwarf galaxy sample}

The surface brightness profiles of dwarf early-type galaxies are usually fitted using a single S\`ersic profile. Few studies in the literature have handled the analysis of the surface brightness distribution of early-type dwarf galaxies using several components. The sample of dwarf galaxies used here comes from: 

\begin{itemize}
\item The sample of Aguerri et al. (2005a). This sample consists of 99 galaxies from the Coma cluster with $-18.0 < M_{B} < -15.7$ and $B-R > 1.25$.  From this sample, 29 objects were classified as two-component dS0s. 
\item The sample of Janz et al. (2014). This sample is formed by 121 galaxies in the Virgo cluster with $-19.0 < M_{r} < -16.0 $. About 70$\%$ of the sample needed more than one structural component in order to fit their 2D surface brightness distributions.
\end{itemize}

The surface brightness profiles of the galaxies from the \cite{aguerri2005a} sample were fitted  using two structural components (S\`ersic + exponential). The S\`ersic profiles were used for the fit of the innermost regions of the galaxies. In addition, an exponential profile was used to fit the surface brightness profile at large radii. In this case, the decision to fit two structural components was made based on the residuals in the surface brightness profiles obtained when only one S\`ersic component was considered in the fits. In this sample, the structural parameters of the S\`ersic and exponential components were obtained in the $R$ band. 

The surface brightness distributions of the sample of dwarf galaxies from \cite{janz2014} were fitted using several structural components. They classify the galaxies into four main groups according to the number of components used in the fits. One-component galaxies turned out to be about 30$\%$ of the sample. The surface brightness distribution of these galaxies was fitted using one S\`ersic profile. Galaxies with two components (S\`ersic + exponential) formed about 40$\%$ of the sample. Three- and four-component galaxies showed bars and/or lenses in addition to the S\`ersic and exponential components. The percentages of galaxies with bars and lenses turned out to be 14$\%$ and 16$\%$  of the galaxies in the sample, respectively. Only dwarf galaxies  with two structural components (S\`ersic + exponential) were considered in the present study. In the case of the  dwarf galaxies in the Virgo cluster, the fit of one or more structural component was decided based on the residuals in the surface brightness profiles of the galaxies and the comparison of the radial trends of the ellipticity and position angle isophotal profiles of the galaxy and the model fitted \citep[see][]{janz2014}. In this sample, the structural parameters of the dS0 galaxies were obtained from images observed throught the $H$ bandpass. 

The sample from \cite{janz2014} is about half a magnitude deeper\footnote{Assuming a colour $g - r \approx 0.8$ for early-type dwarf galaxies, the conversion from $g$ to $B$ bands being taken from \cite{Fukugita96}}. Nevertheless, both galaxy samples cover similar galaxies and contain the bright end of the dwarf galaxy populations in the Virgo and Coma clusters. There are differences in the quality of the images used for the structural decomposition of the dS0 galaxies in the Virgo and Coma clusters. The images of the dS0 galaxies from the Virgo sample have better resolution than those from the Coma cluster owing to the closer distance of the first cluster. In particular, the FWHM of the seeing of the Coma cluster images is 0.65 kpc, while for the Virgo images, it varies between 0.04 and 0.13 kpc. The lower resolution of the Coma cluster images could produce larger uncertainties in the structural parameters of the galaxies, especially for the faintest ones (see Sect. 3.3). Nevertheless, both samples were considered because no other large sample with structural parameters from dS0 galaxies is available in the literature. In addition, detailed galaxy simulations for the Coma cluster galaxies were done in order to obtain the uncertainties of the structural parameters of the galaxies as a function of their apparent magnitude \citep[see][]{aguerri2004}. It was obtained that  the structural parameters of the simulated galaxies were recovered with mean uncertainties smaller than 20\% for those with apparent R-band magnitudes smaller than 17.0. This is the sample of the Coma cluster dS0 galaxies adopted in this study.

\subsection{Filtering the structural parameters}

The structural parameters of the galaxies were filtered in order to obtain a reliable and homogeneous sample of bulge and disc parameters. The first filter was related to the morphological classification of galaxies. The presence of bars or additional structural components affects the bulge and disc structural parameters when only two component models are fitted to the galaxy surface brightness distributions \citep[e.g.\ ][]{aguerri2005c, gadotti2008,salo15}. In particular, \cite{gadotti2008} estimates that the presence of a bar can increase the bulge-to-disc ratio and the effective radius of the bulge by about 20$\%$. Recently, \cite{salo15} has shown that structural decompositions of barred galaxies using two-component models (bulge and disc) overestimate the values of $n$ and the $r_{b,e}/h$ ratio by a large factor. This overestimation is greater in the case of late-type spiral galaxies \citep[see fig. 27 from][]{salo15}. To avoid this effect, only galaxies morphologically classified as spirals (S) or lenticulars (S0) were considered. Galaxies with SB or SB0 morphological classifications were excluded.

The structural parameters of the galaxies were obtained using automatic algorithms. The best solution obtained by these algorithms is a combination of a S\`ersic and  an exponential, thereby minimizing some statistical parameter such as the $\chi^2$.  Nevertheless, the mathematical solution obtained could be a non-physical one.  \cite{allen2006} fitted a large number of surface brightness profiles of galaxies with two structural components and analysed the different types of profiles obtained from automatic fits. They show that all the fitted profiles could be classified into eight different types. The same classification of the fitted profiles was implemented in this study. Finally, only profiles classified as ``classic profiles" \citep[type 1 profiles from the classification in ][]{allen2006} were considered. For these profiles, the S\`ersic component dominates the flux of the galaxy at the innermost region, whereas the disc profile dominates at large radii. In this case the S\`ersic profile represents the bulge component, and the exponential profile models the disc of the galaxy. 

Taking these restrictions into account, the final number of bright and dwarf galaxies used in the present study were 229 and 71, respectively. I notice that 68 bright galaxies were excluded after the filtering proposed here. In most of the cases, the galaxies were excluded because of the morphological type. Only three bright galaxies were excluded only because the fitted structural parameters result in non-physical profiles. 

\subsection{Variation in the structural parameters with wavelength}

\begin{figure}
   \centering
   \includegraphics[width=\hsize]{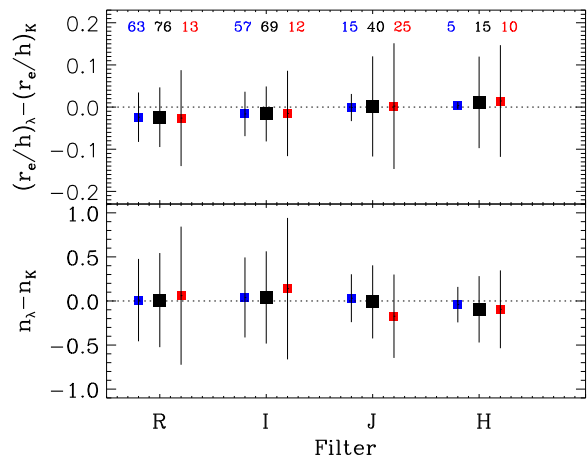}
      \caption{Variation in $r_{b,e}/h$ and $n$ with the bandpass for those galaxies with structural parameters determined in images observed through different wavelengths. Bright galaxies from \cite{graham2003} and \cite{mollenhoff2001} were used. Black squares represent all galaxies. Blue and red squares show galaxies with bulges showing $n < 2$ and $n > 2$, respectively. The numbers located in the top panel indicates the number of total galaxies (black), those with bulges with $n > 2$ (red) and $n < 2$ (blue).
              }
         \label{rehvar}
   \end{figure}

The values of the structural parameters of the galaxies can depend on the wavelength at which they were observed \citep[see e.g.][]{prieto2001}. In the present study, small differences  in the structural parameters of the galaxies due to wavelength variation are expected. This is due to the selection of the bright and dwarf galaxy samples considered here. The structural parameters of the dwarf galaxies were obtained from samples observed in $R$ or $H$ bands. In addition, the structural parameters were obtained from $J$, $H$, or $K$ bands for the bright galaxies. Then, the largest expected variation is between the structural parameters obtained the the $R$ and $K$ bands. To compute and analyse this difference, we studied the variation in the values of $n$ and $r_{b,e}/h$ as a function of the broad band used in the observations. This was done by comparing the structural parameters of the galaxies of the samples observed in several photometric broad-band filters. Figure \ref{rehvar} shows the difference between the values of $n$ and $r_{b,e}/h$ obtained in different bands and those obtained in the$ K$ band. Figure \ref{rehvar} also shows the variation in the structural parameter with wavelength for those galaxies with bulges showing $n > 2$ (red symbols) and $n < 2$ (blue symbols). I would like to point out the small dependence in the structural parameters considered and the wavelength.  We therefore do not expect differences in the results because of the different photometric filters used in observing the  galaxies.

\section{Results}

\subsection{Location of bright galaxies in the $r_{b,e}/h$--$n$ plane}
Figure \ref{rehn} shows the location in the $r_{b,e}/h$--$n$ plane of the bright disc and dS0 galaxies. The bright disc galaxies follow a strong correlation in this plane. The Pearson correlation coefficient of the $r_{b,e}/h$--$n$ relation for bright disc galaxies is 0.58, and the significance of the correlation is greater than 3$\sigma$. The correlation shows that bright disc galaxies with bulges with larger S\`ersic shape parameters also have higher $r_{b,e}/h$ ratio values. 

The relation between the structural parameters of the bulge and disc for bright galaxies has already been analysed in the literature; for example, we may mention the correlation between $r_{b,e}$ and $h$ found in several studies in the past \citep{dejong1996, mollenhoff2001, macarthur2003, mendezabreu2008}. This correlation indicates that larger bulges reside in larger discs, a feature that indicates a common formation process for the bulge and disc components. The relation between $r_{b,e}/h$ and Hubble type\footnote{See \cite{devaucouleurs1976}. The correspondence between the $T$ values and the Hubble morphological types are: $T <-3$ (E), $-3 \leq T \leq-1$ (S0), $T=0$ (S0/a), $T=1$ (Sa), $T=3$ (Sb), $T=5$ (Sc), $T=7$ (Sd), $T=9$ (Sm), and $T=10$ (Irr).} has also been analysed in the past. Several results indicate that $r_{b,e}/h$ does not depend on the Hubble type \citep{dejong1996, balcells2007, graham2008}. In this case, this means that the Hubble sequence is scale-free, and the bulge-disc size ratio cannot be used for the morphological classification of disc galaxies. In contrast, other studies show some variation in the $r_{b,e}/h$ ratio with the Hubble type \citep[see][]{graham1999, graham2001}. The $r_{b,e}/h$ vs $n$ relation presented in Fig. \ref{rehn} has also been reported in the literature \citep[e.g.\  ][]{mendezabreu2008, debattista2006}. Numerical simulations of bulges and discs of galaxies formed in hierarchical clustering scenarios report similar relations between the ratio of the bulge and disc scale lengths and the S\`ersic shape parameter of the bulge \citep[see][]{scannapieco2003, tissera2006}.

To understand the physics behind the obtained relation between $r_{e,b}/h$ and $n$ for our sample of bright galaxies, we analysed the dependence of these two quantities on other parameters, such as Hubble type ($T$), bulge absolute magnitude in the K-band $(M_{B,K})$, bulge-to-total ratio ($B/T$), and disc K-band absolute magnitude ($M_{D,K}$). Figure \ref{rehnvariation} shows these relations. All the quantities were obtained from the structural decompositions of the bright disc galaxies considered in the present work. The Pearson test shows that the significance of the correlations between $r_{e,b}/h$ and $T$, $M_{B,K}$, and $B/T$ are greater than 3$\sigma$. In contrast, this test reports no significant correlation between $r_{e,b}/h$ and $M_{D,K}$. The strongest correlations are $r_{e,b}/h$ vs $T$ and $r_{e,b}/h$ vs $B/T$. The dependence of $n$ is evident with all galaxy parameters we studied. The Pearson test shows that the significance of the correlation between $n$ and the other parameters are greater than 3$\sigma$ for all cases. These relations indicate that those bright galaxies selected in the present work with higher $r_{e,b}/h$ values are those with prominent bulges and higher values of $n$. The prominence of the bulge is one of the main physical parameters driving the Hubble galaxy classification. The correlation between $r_{e,b}/h$ and $n$ for bright galaxies is therefore a consequence of the prominence of the bulge and therefore of the Hubble sequence.

\subsection{Location of dS0 galaxies in the $r_{b,e}/h$--$n$ plane}

\begin{figure*}
   \centering
   \includegraphics[width=18.cm]{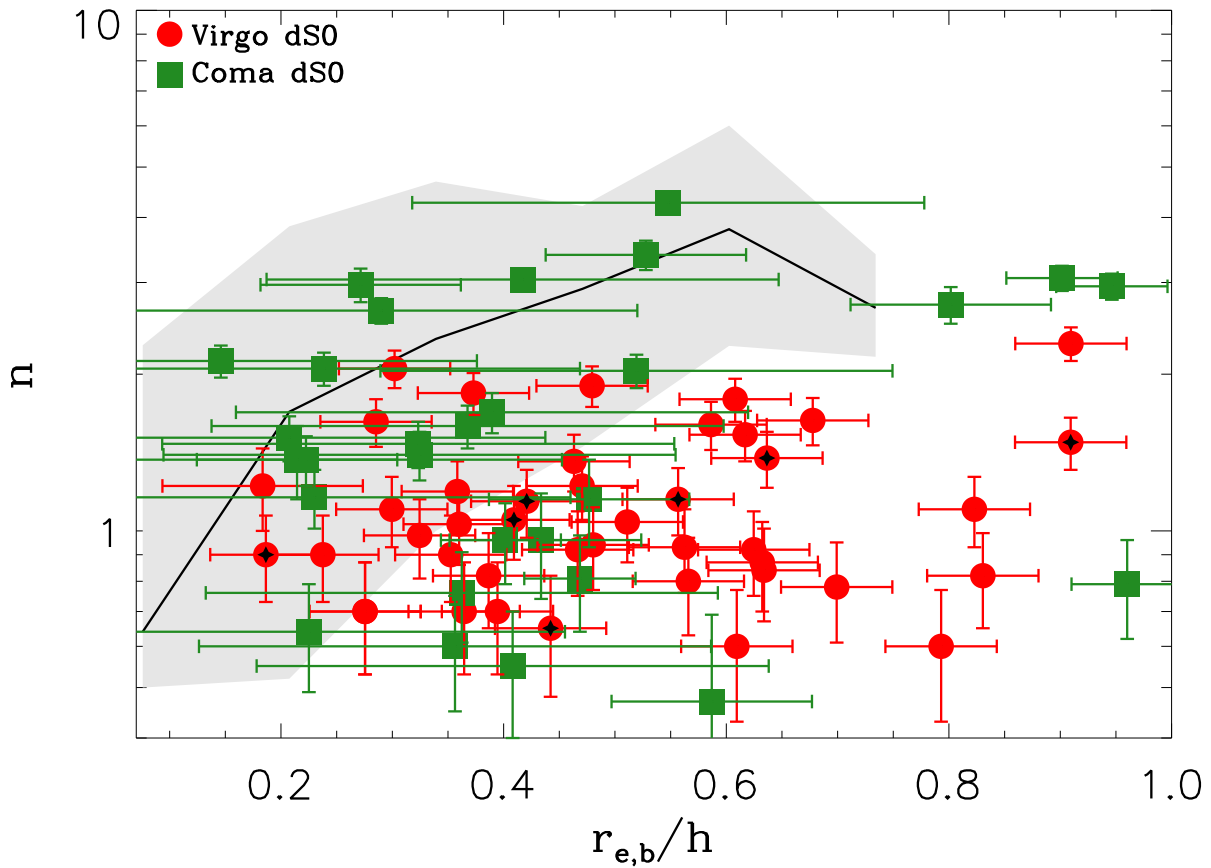}
      \caption{Distribution of bright disc galaxies (grey region) and dS0 (red dots and green squares) in the $r_{b,e}/h$--$n$ plane.  The grey region represents the locus of 95$\%$ of the bright disc galaxies in this plane. The solid line shows the median of the distribution of the bright disc galaxies.  The upper right corner of the $r_{b,e}/h$--$n$ relation for bright disc galaxies is mainly made by bulge-dominated systems. In contrast, disc dominated systems are mainly located in the lower left corner of the relation. The black stars represent fast rotators according to the classification given by \cite{toloba14}. 
              }
         \label{rehn}
   \end{figure*}

\begin{figure*}
   \centering
   \includegraphics[width=15.0cm,angle=90]{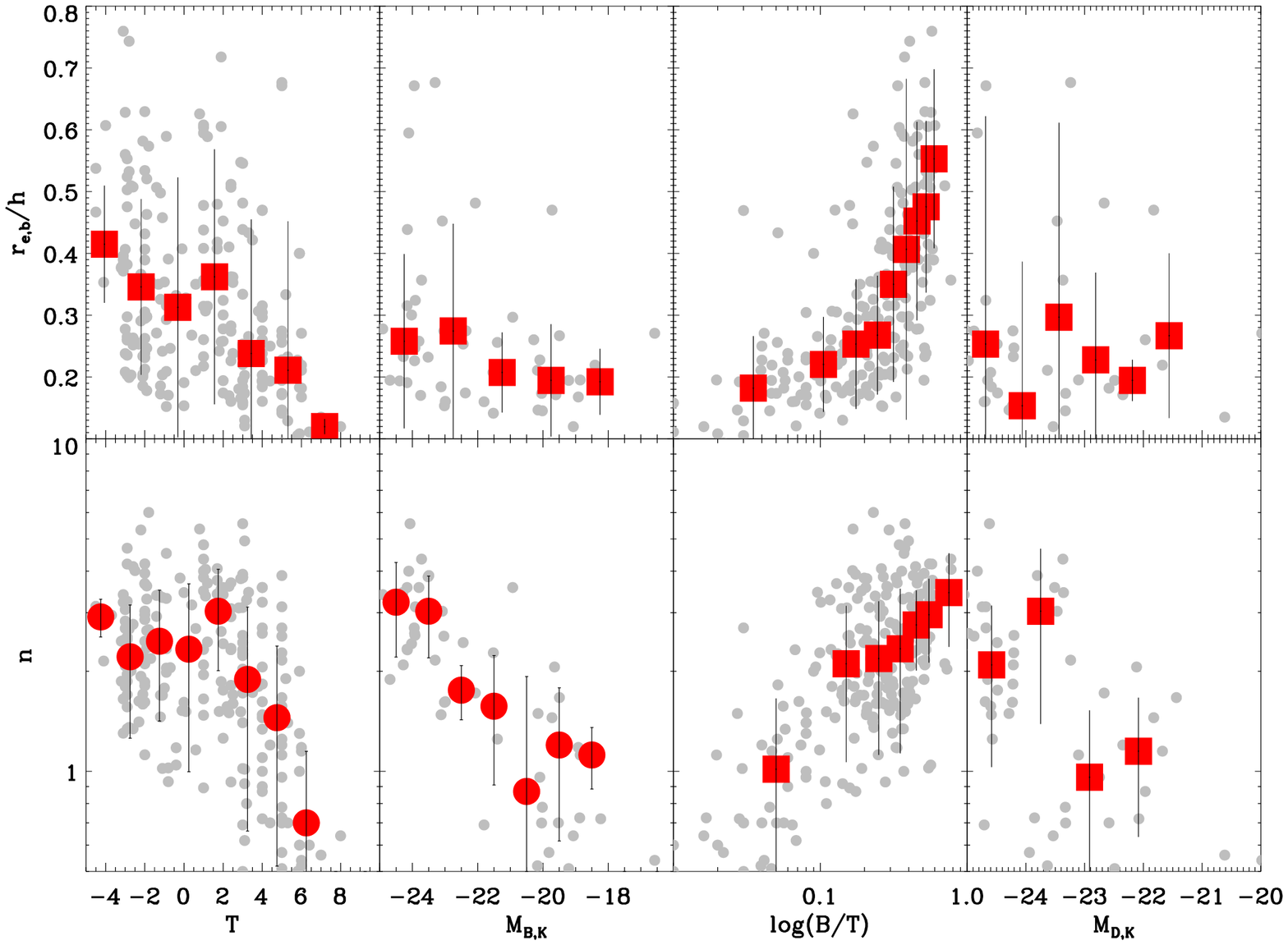}
      \caption{Dependence of $r_{e,b}/h$ and $n$ with Hubble type ($T$), K-band bulge absolute magnitude ($M_{b,K}$), bulge-to-total ratio ($B/T$), and K-band disc absolute magnitude ($M_{D,K}$). Grey points represent the data considered in the present paper. The red points show the mean values of $r_{e,b}/h$ and $n$ as a function of the different galactic parameters. The errors represent the dispersion.
              }
         \label{rehnvariation}
   \end{figure*}

The distribution of dS0 galaxies in the $r_{b,e}/h$--$n$ plane differs from that for bright disc galaxies. Figure \ref{rehn} shows that dS0s do not follow any relation in the $r_{b,e}/h$--$n$ plane. In addition, a fraction of dS0 galaxies ($\sim$$55\%$) are located outside the locus occupied by 95$\%$ of bright disc galaxies\footnote{This percentage was computed by considering only those dS0 galaxies having values of $r_{e,b}/h$ and $n$ in the same range as those from bright disc galaxies.}. The statistical differences in the distribution of bright and dS0 galaxies have also been computed by using the two-dimensional Kolmogorov-Smirnov (KS2D) test. This test gives the probability that two distribution of points come from the same distribution function. The results of the KS2D test on the points represented in Fig. \ref{rehn} returns than bright and dwarf galaxies are statistically different  at significance level better than 99.9$\%$. This differences are also obtained when only bright and dS0 galaxies from the Virgo and the Coma clusters are considered separately. 

The dS0 galaxies show low values of $n$. In particular, all dS0 from the Virgo cluster and 70$\%$ of those from the Coma cluster have $n < 2$. The low values of $n$ of the dS0 galaxies are expected because the $n$ parameter strongly correlates with the luminosity or mass of the spheroid component of the galaxies \citep[see e.g.][]{trujillo2002}. Nevertheless, dS0 galaxies show higher values of $r_{b,e}/h$ than the expected ones for the low values of $n$ that they have.  The previous section shows that the parameter $r_{b,e}/h$ in the bright disc galaxies strongly correlates with T or B/T parameters. Thus, dS0 galaxies with large $r_{b,e}/h$ could indicate that they have more prominent bulges than those expected for the low $n$ values that they show. This is evident after the analysis of the $B/T$ values of bright discs and dS0 galaxies showing $n<2$. The median $B/T$ values of bright and dS0 galaxies are 0.17 and 0.21, respectively. The KS test shows that the probability distribution functions of $B/T$ for bright discs and dS0 galaxies with $n<2$ are statistically different at 95$\%$ C.L. In the case of dS0 galaxies outside from the locus described by the bright discs in the $r_{b,e}/h$--$n$ plane, the differences in $B/T$ are greater. In this case, the median $B/T$ is 0.24 and the two $B/T$ families come from different distribution functions at 99$\%$ C.L.

Figure \ref{rehn} shows some differences between the dS0 galaxies in the Coma and Virgo clusters.  In particular, the number of  dS0 galaxies located in the locus defined by bright disc galaxies in the $r_{b,e}/h$ -- $n$ plane is larger in the Coma cluster than in Virgo. One possible explanation could be that the lower mass of the Virgo cluster  favours tidal interactions. This would produce greater mass loss for the dS0 galaxies in the Virgo cluster and would result in large $r_{b,e}/h$ ratios. Nevertheless, the quality of the data and the number of galaxies studied cannot fully support this statement. Thus, the better resolution of the bulges of the Virgo galaxies, along with systematics in the selection of the samples or in the computation of their structural parameters, could also produce the observed differences.  New data from a larger sample of nearby clusters would be needed to confirm the observed differences in the structural parameters of the dS0 galaxies of the two clusters.

\subsection{Uncertainties in $r_{b,e}/h$ and $n$ for dwarf galaxies}

Figure 2 shows the uncertainties in $r_{b,e}/h$ and $n$ structural parameters of the Virgo and Coma galaxies. These uncertainties were obtained by using the Monte Carlo simulations from \cite{aguerri2004}. In these simulations, galaxies with similar sizes and luminosities to those expected in the Coma cluster were simulated in order to evaluate the uncertainties in the fit of their structural parameters. In particular, we selected those simulated galaxies with $17 < m_{r} < 19$. This set of galaxies contains dwarf galaxies down to $M_{r} < -16$ at the distance of the Coma cluster. The selected simulated galaxies were grouped into three families according to the $r_{b,e}/PSF$ ratio, where $PSF$ represents the FWHM of the point-spread function of the images of the Coma cluster. The three groups of simulated galaxies where those with $r_{b,e}/PSF < 1.2$, $1.2 < r_{b,e}/PSF < 1.5$ and $r_{b,e}/PSF > 1.5$. Finally, the mean absolute uncertainties of $r_{b,e}/h$ and $n$ were computed for the three groups of simulated galaxies. As expected, larger uncertainties were obtained in the structural parameters of less resolved galaxies.

Both the Virgo and the Coma dS0 observed galaxies were also classified according to their $r_{b,e}/PSF$ ratio. The assigned uncertainties to $r_{e,b}/h$ and $n$ were  those computed using the simulated galaxies. These are the uncertainties shown in Fig. 2 for the Virgo and Coma dS0 galaxies. Larger uncertainties of $r_{e,b}/h$ parameter are shown by the dS0 galaxies in the Coma cluster. This is because these galaxies are less resolved than those in the Virgo cluster. Considering these uncertainties, we can conclude that $\sim40\%$ of the dS0 galaxies are located outside the locus occupied by 95$\%$ of the bright galaxies.

\section{Discussion}

According to the distribution of dS0 galaxies in the $r_{b,e}/h$--$n$ plane, we can say that dS0 galaxies are not a low mass replica of bright disc galaxies. In contrast, some of the dS0 galaxies are located in a region of the plane similar to the one occupied by tidally truncated disc galaxies. This distribution could indicate that the progenitors of these dS0 galaxies are bright or massive disc galaxies tidally truncated by the cluster environment. 

Numerical simulations show that the remnants of tidally truncated disc galaxies carry clues to their tidal origin \citep{aguerri2009}. In particular, rotation in the external regions of galaxies can be expected in the case of dS0 galaxies being tidally truncated disc galaxies. In the present section, I discuss this situation by analysing  the location of the fast rotators in the $r_{b,e}/h$--$n$ plane. I also analyse the distribution of dS0 galaxies in the phase space of clusters. This approach will provide information about the infall time of dS0 galaxies into the clusters. I also study the dependence of the results on the selection of the galaxy samples. Finally, I discuss other scenarios for the formation of dS0 galaxies.

\subsection{Selection effects}

The structural parameters of the bright galaxies used in the present study were obtained from various samples in the literature. These samples used different algorithms for computing the structural parameters. In addition, they also have different selection criteria for the galaxies analysed.  The main result presented here is based on the strong correlation between $r_{b,e}/h$ and $n$ for the bright galaxies. Could this relation be a product of a selection bias in the bright galaxy sample?

To study the dependence of these inhomogeneities in the distribution of bright galaxies in the $r_{b,e}/h$--$n$ plane, the distribution of the disc and massive galaxies presented in the Spitzer Survey of Stellar Structure in Galaxies \citep[$S^{4}G$, ][]{sheth10} were analysed. This is a deep 3.6 and 4.5 $\mu m$ imaging survey of nearby galaxies. These galaxies have morphological classifications taken from HyperLEDA\footnote{I acknowledgeusing the HyperLEDA database (http://leda.univ-lyon1.fr)}. The structural parameters of these galaxies were presented in \cite{salo15}. They fitted the two-dimensional surface brightness distribution of the galaxies by multicomponent models including bulge, disc, bars, and other components. The selection of the galaxies in this sample is homogenous, and their structural parameters were obtained using the same numerical algorithm. I filtered the structural parameters of the galaxies in a similar way as described in Section 2.3. I considered only galaxies with two components (bulge and disc). In addition, the resulting fitted profiles can be classified as classic profiles in the sense that the bulge structures dominate the surface brightness in the inner regions and the disc in the outermost ones.

\begin{figure}
   \centering
   \includegraphics[width=9.3cm]{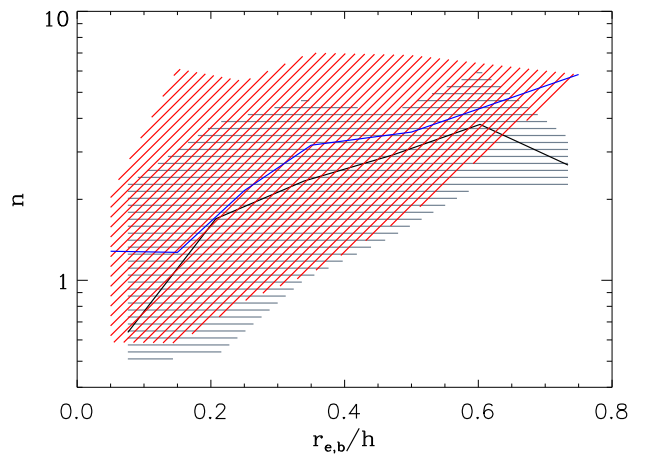}
      \caption{Distribution of bright/massive galaxies in the $r_{b,e}/h$--$n$ plane. The shaded areas represent the locus of 95$\%$ of the galaxies from: galaxies with $M > 10^{10} M_{\odot}$ from Salo et al. (2015) (red shaded region) and bright galaxies used in the present study (grey shaded region). The blue and black full lines represent the position of 50$\%$ of the galaxies in the Salo et al (2015) and the bright galaxies of the present study, respectively.
              }
         \label{s4g}
   \end{figure}

The catalogue of \cite{munozmateos13} provides information on the stellar masses of the $S^{4}G$ galaxies. I selected galaxies with $M > 10^{10} M_{\odot}$ in order to analyse their distribution in the $r_{b,e}/h$--$n$ plane (see Fig. \ref{s4g}). The relation between $r_{b,e}/h$ and $n$ for these massive galaxies from the $S^{4}G$ sample is similar to the relation shown by the bright galaxy sample used in this study. In particular, it is observed that galaxies with higher $r_{b,e}/h$ values show larger $n$ parameters. In addition, the lower limit of the locus of 95$\%$ of the massive $S^{4}G$ galaxies is similar to the limit defined by the bright galaxy sample used in this work. 

I may therefore conclude that no bias in the results presented in this work is  expected owing to sample selection effects.

%

\subsection{One possible formation scenario of dS0 galaxies}

Section 3  shows that bright and dS0 galaxies do not follow the same correlation in the $r_{e,b}/h$ vs $n$ plane. In particular, about 50$\%$ of the dS0 galaxies are located out from the locus defined by the 95$\%$ of the bright galaxies in this plane. Do these differences indicate something about the formation history of dS0 galaxies? In particular, can dS0 galaxies be remnants of strongly transformed bright disc galaxies by tidal interactions in the clusters?

Aguerri \& Gonz\'alez-Garc\'{\i}a (2009) produced a set of numerical simulations in order to show the effects of tidal fast encounters in bulge-disc galaxies. In particular, these simulations analysed the structure and kinematics of a bulge-disc galaxy after several fast tidal interactions similar to those suffered by galaxies in cluster potentials. The initial conditions considered massive galaxies ($8\times 10^{11} - 1.1 \times 10^{11} M_{\odot}$) with different bulge-to-disc ratios ($0.01 < B/D < 0.5$). In all cases, the initial values of $n$ for the bulge component of the galaxies were about 1.0. 

The simulations show that after several fast interactions, the galaxies lose a large amount of their mass. Tidal disruption mainly affects the external regions of the galaxies. Both the dark-matter halo and the disc of the galaxies are strongly truncated. The stellar discs of the galaxies have reduced their scale lengths by about 30--50\%. The bulge component also expands and increases its $r_{e,b}$ values. The mean values $r_{e,b}/h$ in the models after several fast interactions are between 0.1-0.6 more than the initial ones. These interactions therefore produce a sharp change in the values of $r_{e,b}/h$ of the galaxies. In contrast, no significant change in the values of $n$ are observed in the models \citep[see fig. 15 in][]{aguerri2009}

This tidal interaction scenario could explain the observed differences between bright and dS0 galaxies in the $r_{e,b}/h$-$n$ plane. Thus, tidal interactions could transform bright disc galaxies in dwarf-like systems. The resulting remnant would have similar values of $n$ as the progenitor, but higher values of $r_{e,b}/h$. 

The simulations from \cite{aguerri2009} can explain the dS0 galaxies with $n$ about 1. More simulations with higher initial values of $n$ should be run in the future to properly explain all the dS0 galaxies.

\cite{janz2014} find that, at a given magnitude, the bulge components of the dS0 galaxies in the Virgo cluster are systematically larger than bulges of bright spirals. They conclude that this difference in the scales point towards the inner structural component of the dS0 galaxies not being ordinary bulges. It was mentioned above that tidal interactions produce an expansion of the bulge component of the galaxies \citep[see][]{aguerri2009}. Then, larger spheroidal components in the dS0 galaxies than the bulges of spirals are expected in the tidal scenario. Nevertheless, numerical interactions also indicate that the mass of the bulge component of the tidally truncated disc galaxies does not change \citep[see][]{aguerri2009}. This means that it would be the mass of the inner component of the dS0 and the mass of bulges of bright spiral galaxies that should be compared.

\subsection{Are dS0 galaxies fast-rotator systems?}

Simulations also show that tidally truncated bright disc galaxies conserve some memory of their initial dynamical state. Although several fast tidal interactions can disrupt a large portion of their mass, the remnants still maintain rotation in their outermost regions \citep[see][]{aguerri2009}.  This property implies that we should expect some rotation at large radii in dS0 galaxies if they are remnants of tidally truncated disc galaxies. To see whether the dS0 population has some indication of rotation in their external regions, I examined their stellar kinematical properties in the literature.

\cite{toloba14} studied the stellar kinematics of 39 early-type dwarf galaxies in the Virgo cluster. They classified these galaxies in fast and slow rotators based on their specific angular momentum and ellipticity. We have identified nine galaxies in common between this sample and that of  \cite{janz2014} used in the present study. Seven of these common galaxies were classified as fast rotators by \cite{toloba14}.  Six of the fast-rotator galaxies are located outside the locus defined by 95$\%$ of the bright disc galaxies in the $r_{b,e}/h$ vs $n$ plane. This means that most of them show some indication of rotation in their external regions, as expected if their progenitors were bright disc galaxies.

   \begin{figure}
   \centering
   \includegraphics[width=9.3cm]{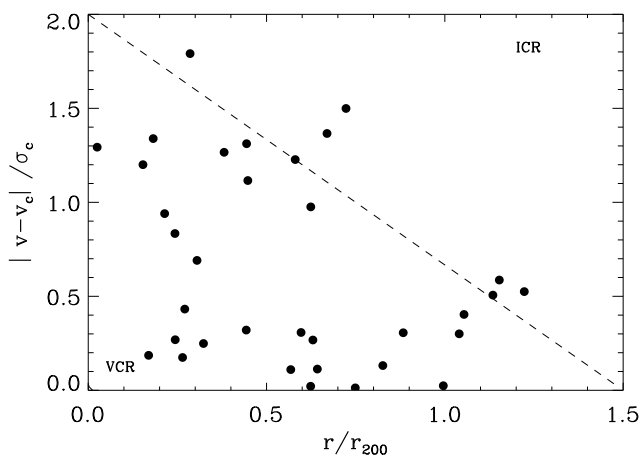}
      \caption{Distribution of dS0 galaxies (black points) in the phase-space diagram. The diagonal dashed line divides the phase-space diagram between the infall cluster region (ICR) and the virialized cluster region (VCR) \citep[see][]{oman13}. 
              }
         \label{caustic}
   \end{figure}

\subsection{Are dS0 galaxies recent arrivals to the cluster?}

Galaxies in clusters can evolve through the action of several physical mechanisms that can affect different galaxy components. In addition, the action of the physical mechanisms can be stronger in specific parts of the clusters. For example, ram-pressure stripping is stronger in the densest cluster regions where the gas density and the velocity of the galaxies are higher. Tidal interactions with the cluster potential are expected to be stronger in the central cluster regions when the galaxies pass through their orbital pericentre. Nevertheless, fast encounters with other galaxies in the cluster can take place at large cluster-centric distances. The time scales of these mechanisms are also different and can vary from a few Myr for ram-pressure stripping to a few Gyr for harassment. All these physical mechanisms and time scales make disentangling, which is the dominant mechanism driving galaxy evolution in clusters, a difficult challenge. It is probable that galaxies are transformed by several mechanisms during their life in the cluster.

We may wonder about the time of arrival of the dS0 galaxies studied here at the Virgo and the Coma clusters. In particular, we may distinguish between galaxies that have recently arrived at the cluster and those that have spent several Gyr living inside it by studying the position of galaxies in the phase-space diagram. Numerical simulations show that galaxy haloes located in the innermost regions of the cluster arrive, on average, earlier than those located in the outermost regions \citep[see][]{haines12}. The position of the galaxies in phase space can also tell us about the arrival time at the cluster. Recently, \cite{oman13} computed the probability distribution functions of the time of arrival at the cluster of haloes of galaxies located at different positions in the phase-space diagram defined by $|$v-v$_{c}|/\sigma_{c}$ vs $r/r_{200}$, where v is the line-of-sight velocity of the halo, v$_{c}$ and $\sigma_{c}$ are the mean velocity and velocity dispersion of the cluster, and $r$ and $r_{200}$ are the radial distance of the galaxy to the cluster centre and the galaxy cluster virial radius. They defined a line in the phase-space splitting infalling galaxies and those residing in the cluster during several Gyr. \citep[see fig. 4 from ][]{oman13}

Figure \ref{caustic} shows the distribution of the dS0 galaxies used in the present study in the phase-space diagram. The  dS0  galaxies are spread throughout the phase space. Nevertheless, a very small fraction of them can be considered as infalling galaxies,  according to the \cite{oman13} result. Most of the dS0 galaxies arrived at the cluster potential several Gyr ago and have had enough time to undergo tidal interactions with the cluster potential or the cluster substructure.

\subsection{Other possible formation scenarios of dS0 galaxies}

The tidal formation scenario for dS0 galaxies is the preferred one in the present paper. Nevertheless, there are several physical processes that galaxies suffer in high density environments that can influence the formation and evolution of dwarf systems. In particular, those physical mechanisms related to the gas content of the galaxies can influence the morphology and the structural parameters of dwarf and bright galaxies. Thus, ram-pressure stripping \citep[see][]{gunn1972, quilis2000, bekki2009}, strangulation \citep[see][]{kawata2008}, and starvation \citep[see][]{larson1980, bekki2002} produce the sweep of the cold and/or hot gas from the galaxies. This loss of gas stops the active star formation of the galaxies and produces a decrease in their stellar surface brightness \citep[see e.g.][]{boselli2014,boselli2014b, rodriguezdelpino2014}. In particular, the end of the star formation in the outer galaxy regions would produce a fading of these regions, producing shorter length scales of the discs of the galaxies.  This would produce a change in the $r_{b,e}/h$ ratio in a similar way to tidal interactions. The variation in the structural parameters would be more pronounced when the galaxies are observed through optical bands, so more influenced by the light of young and intermediate age stars. 

Some differences could be expected between the tidal and the gas loss scenarios. In particular, the external regions of the galaxies are completely truncated in the former scenario \citep[see e.g.][]{mastropietro2005,aguerri2009, bialas2015}. The stars and gas previously located in the external regions of the galaxies are disrupted by the interaction and located in the intracluster region. In contrast, in the gas loss scenarios, the stars in the external regions of the galaxies are not swept out \citep[see e.g.][]{fossati2013}. Nevertheless, these external regions should have very low surface brightness, which makes them difficult to observe. The presence or absence of truncation in the external regions of the galaxies must also appear in the kinematics of the galaxies \citep[see e.g.][]{jaffe2011}. Thus, deep studies of the external regions of the dS0 galaxies could help us to differentiate between tidal and gas-stripping scenarios. In addition, detailed simulations of how the swept of the gas affects to the external parts of the galaxies and its structural parameters would help to distinguish between these scenarios.

\section{Conclusions}

The present study has analysed the location of a sample of bright disc galaxies in the $r_{b,e}/h$--$n$ plane. Bright disc galaxies show a strong correlation between $r_{b,e}/h$ and $n$. In particular, galaxies with bulges showing higher values of the S\`ersic $n$ parameter also have higher $r_{b,e}/h$ ratio values. In contrast, multicomponent early-type dwarf galaxies do not follow this correlation. They are spread throughout the $r_{b,e}/h$--$n$ plane. In addition, a large number of them ($\sim$$55\%$) are located outside the locus defined by 95$\%$ of the bright disc galaxies in the $r_{b,e}/h$--$n$ plane. These dS0 galaxies are located in the same position as the remnants of strong tidally truncated disc galaxies, as described by high resolution numerical simulations. Therefore, these dS0 galaxies in clusters would be the end products of  galaxies strongly transformed by tidal interactions. This galaxy population represents about 20-25$\%$ of all early-type dwarf galaxies with $-18 \leq M_{B} \leq -15$ in the Virgo and the Coma clusters.

This study shows that the $r_{b,e}/h$--$n$ plane can be used to select candidates for end products of  galaxies in clusters strongly transformed  by environmental effects. Future detailed studies of the stellar kinematics and/or the stellar populations of these galaxies can provide additional information on the origin and evolution of this galaxy cluster population.

\begin{acknowledgements}
      This work was partially supported by the Spanish Ministerio de Econom\'{\i}a y Competitividad through project AYA2013-43188-P. The author also thanks R. Smith, S. Courtea, S. Zarattini, and T. Lisker for useful discussions during the preparation of this manuscript.
\end{acknowledgements}



\begin{thebibliography}{}

\bibitem[Aguerri \& Gonz{\'a}lez-Garc{\'{\i}}a(2009)]{aguerri2009} Aguerri, J.~A.~L., \& Gonz{\'a}lez-Garc{\'{\i}}a, A.~C.\ 2009, \aap, 494, 891 
\bibitem[Aguerri et al.(2005a)]{aguerri2005a} Aguerri, J.~A.~L., Iglesias-P{\'a}ramo, J., V{\'{\i}}lchez, J.~M., Mu{\~n}oz-Tu{\~n}{\'o}n, C., \& S{\'a}nchez-Janssen, R.\ 2005, \aj, 130, 475
\bibitem[Aguerri et al.(2005b)]{aguerri05b} Aguerri, J.~A.~L., 
Gerhard, O.~E., Arnaboldi, M., et al.\ 2005, \aj, 129, 2585 
\bibitem[Aguerri et al.(2005c)]{aguerri2005c} Aguerri, J.~A.~L., Elias-Rosa, N., Corsini, E.~M., \& Mu{\~n}oz-Tu{\~n}{\'o}n, C.\ 2005, \aap, 434, 109 
\bibitem[Aguerri et al.(2004)]{aguerri2004} Aguerri, J.~A.~L., Iglesias-Paramo, J., Vilchez, J.~M., \& Mu{\~n}oz-Tu{\~n}{\'o}n, C.\ 2004, \aj, 127, 1344 
\bibitem[Agulli et al.(2014)]{agulli2014} Agulli, I., Aguerri, J.~A.~L., S{\'a}nchez-Janssen, R., et al.\ 2014, \mnras, 444, L34 
\bibitem[Allen et al.(2006)]{allen2006} Allen, P.~D., Driver, S.~P., Graham, A.~W., et al.\ 2006, \mnras, 371, 2 
\bibitem[Amor{\'{\i}}n et al.(2009)]{amorin09} Amor{\'{\i}}n, R., Aguerri, J.~A.~L., Mu{\~n}oz-Tu{\~n}{\'o}n, C., \& Cair{\'o}s, L.~M.\ 2009, \aap, 501, 75 
\bibitem[Andredakis et al.(1995)]{andredakis1995} Andredakis, Y.~C., Peletier, R.~F., \& Balcells, M.\ 1995, \mnras, 275, 874 
\bibitem[Arnaboldi et al.(2002)]{arnaboldi02} Arnaboldi, M., Aguerri, J.~A.~L., Napolitano, N.~R., et al.\ 2002, \aj, 123, 760 

\bibitem[Balcells et al.(2007)]{balcells2007} Balcells, M., Graham, A.~W., \& Peletier, R.~F.\ 2007, \apj, 665, 1104 
\bibitem[Barazza et al.(2003)]{barazza03} Barazza, F.~D., Binggeli, B., \& Jerjen, H.\ 2003, \aap, 407, 121 
\bibitem[Bekki(2009)]{bekki2009} Bekki, K.\ 2009, \mnras, 399, 2221 
\bibitem[Bekki et al.(2002)]{bekki2002} Bekki, K., Couch, W.~J., \& Shioya, Y.\ 2002, \apj, 577, 651 
\bibitem[Benson(2014)]{benson2014} Benson, A.~J.\ 2014, \mnras, 444, 2599 
\bibitem[Binggeli \& Cameron(1991)]{binggeli91} Binggeli, B., \& Cameron, L.~M.\ 1991, \aap, 252, 27
\bibitem[Binggeli et al.(1984a)]{binggeli84a} Binggeli, B., Sandage, A., \& Tarenghi, M.\ 1984, \aj, 89, 64
\bibitem[Sandage \& Binggeli(1984b)]{binggeli84b} Sandage, A., \& Binggeli, B.\ 1984, \aj, 89, 919 
\bibitem[Bialas et al.(2015)]{bialas2015} Bialas, D., Lisker, T., Olczak, C., Spurzem, R., \& Kotulla, R.\ 2015, \aap, 576, A103 
\bibitem[Blanton et al.(2005)]{blanton2005} Blanton, M.~R., Lupton, R.~H., Schlegel, D.~J., et al.\ 2005, \apj, 631, 208 
\bibitem[Boselli \& Gavazzi(2014)]{boselli2014} Boselli, A., \& Gavazzi, G.\ 2014, \aapr, 22, 74 
\bibitem[Boselli et al.(2014)]{boselli2014b} Boselli, A., Voyer, E., Boissier, S., et al.\ 2014, \aap, 570, A69
\bibitem[Boselli et al.(2011)]{boselli2011} Boselli, A., Boissier, S., Heinis, S., et al.\ 2011, \aap, 528, A107
\bibitem[Boselli et al.(2008)]{boselli2008} Boselli, A., Boissier, S., Cortese, L., \& Gavazzi, G.\ 2008, \apj, 674, 742 
\bibitem[Boselli \& Gavazzi(2006)]{boselli2006} Boselli, A., \& Gavazzi, G.\ 2006, \pasp, 118, 517 


\bibitem[Caon et al.(1993)]{caon93} Caon, N., Capaccioli, M., \& D'Onofrio, M.\ 1993, \mnras, 265, 1013
\bibitem[Castro-Rodrigu{\'e}z et al.(2009)]{castrorodriguez09} Castro-Rodrigu{\'e}z, N., Arnaboldi, M., Aguerri, J.~A.~L., et al.\ 2009, \aap, 507, 621  
\bibitem[C{\^o}t{\'e} et al.(2006)]{cote2006} C{\^o}t{\'e}, P., Piatek, S., Ferrarese, L., et al.\ 2006, \apjs, 165, 57 

\bibitem[Debattista et al.(2006)]{debattista2006} Debattista, V.~P., Mayer, L., Carollo, C.~M., et al.\ 2006, \apj, 645, 209 
\bibitem[de Jong(1996)]{dejong1996} de Jong, R.~S.\ 1996, \aap, 313, 45 
\bibitem[de Jong \& van der Kruit(1994)]{dejong1994} de Jong, R.~S., \& van der Kruit, P.~C.\ 1994, \aaps, 106, 451 
\bibitem[De Propris et al.(2003)]{depropris2003} De Propris, R., Colless, M., Driver, S.~P., et al.\ 2003, \mnras, 342, 725 
\bibitem[de Vaucouleurs et al.(1976)]{devaucouleurs1976} de Vaucouleurs, G., de Vaucouleurs, A., 
\& Corwin, J.~R.\ 1976, Second reference catalogue of bright galaxies, Vol.~1976, p.~Austin: University of Texas Press., 1976,  

\bibitem[Fossati et al.(2013)]{fossati2013} Fossati, M., Gavazzi, G., Savorgnan, G., et al.\ 2013, \aap, 553, A91 
\bibitem[Freeman(1970)]{freeman1970} Freeman, K.~C.\ 1970, \apj, 160, 811 
\bibitem[Fukugita et al.(1996)]{Fukugita96} Fukugita, M., Ichikawa, T., Gunn, J.~E., et al.\ 1996, \aj, 111, 1748 

\bibitem[Gadotti(2008)]{gadotti2008} Gadotti, D.~A.\ 2008, \mnras, 384, 420 
\bibitem[Gavazzi et al.(2013)]{gavazzi2013} Gavazzi, G., Fumagalli, M., Fossati, M., et al.\ 2013, \aap, 553, A89 
\bibitem[Gavazzi et al.(2010)]{gavazzi2010} Gavazzi, G., Fumagalli, M., Cucciati, O., \& Boselli, A.\ 2010, \aap, 517, A73 
\bibitem[Gavazzi et al.(2005)]{gavazzi2005} Gavazzi, G., Donati, A., Cucciati, O., et al.\ 2005, \aap, 430, 411
\bibitem[Geha et al.(2003)]{geha2003} Geha, M., Guhathakurta, P., \& van der Marel, R.~P.\ 2003, \aj, 126, 1794
\bibitem[Gerhard et al.(2005)]{gerhard05} Gerhard, O., Arnaboldi, M., Freeman, K.~C., et al.\ 2005, \apjl, 621, L93 
\bibitem[Graham(2001)]{graham2001} Graham, A.~W.\ 2001, \aj, 121, 820 
\bibitem[Graham \& Prieto(1999)]{graham1999} Graham, A.~W., \& Prieto, M.\ 1999, \apjl, 524, L23 
\bibitem[Graham \& Worley(2008)]{graham2008} Graham, A.~W., \& Worley, C.~C.\ 2008, \mnras, 388, 1708 
\bibitem[Graham(2003)]{graham2003} Graham, A.~W.\ 2003, \aj, 125, 3398 
\bibitem[Gunn \& Gott(1972)]{gunn1972} Gunn, J.~E., \& Gott, J.~R., III 1972, \apj, 176, 1 

\bibitem[Haines et al.(2012)]{haines12} Haines, C.~P., Pereira, M.~J., Sanderson, A.~J.~R., et al.\ 2012, \apj, 754, 97 
\bibitem[Haines et al.(2006)]{haines2006} Haines, C.~P., La Barbera, F., Mercurio, A., Merluzzi, P., \& Busarello, G.\ 2006, \apjl, 647, L21 
\bibitem[Hogg et al.(2004)]{hogg2004} Hogg, D.~W., Blanton, M.~R., Brinchmann, J., et al.\ 2004, \apjl, 601, L29 

\bibitem[Jaff{\'e} et al.(2011)]{jaffe2011} Jaff{\'e}, Y.~L., Arag{\'o}n-Salamanca, A., Kuntschner, H., et al.\ 2011, \mnras, 417, 1996 
\bibitem[Janz et al.(2014)]{janz2014} Janz, J., Laurikainen, E., Lisker, T., et al.\ 2014, \apj, 786, 105 
\bibitem[Jerjen \& Binggeli(1997)]{jerjen1997} Jerjen, H., \& Binggeli, B.\ 1997, The Nature of Elliptical Galaxies; 2nd Stromlo Symposium, 116, 239 


\bibitem[Kawata \& Mulchaey(2008)]{kawata2008} Kawata, D., \& Mulchaey, J.~S.\ 2008, \apjl, 672, L103 
\bibitem[Kenney et al.(2004)]{kenney2004} Kenney, J.~D.~P., van Gorkom, J.~H., \& Vollmer, B.\ 2004, \aj, 127, 3361 
\bibitem[Koleva et al.(2013)]{koleva2013} Koleva, M., Bouchard, A., Prugniel, P., De Rijcke, S., \& Vauglin, I.\ 2013, \mnras, 428, 2949 
\bibitem[Kormendy(1985)]{kormendy1985} Kormendy, J.\ 1985, \apj, 295, 73 
\bibitem[Kronberger et al.(2008)]{kronberger2008} Kronberger, T., Kapferer, W., Unterguggenberger, S., Schindler, S., \& Ziegler, B.~L.\ 2008, \aap, 483, 783 
\bibitem[Kormendy et al.(2009)]{kormendy09} Kormendy, J., Fisher, D.~B., Cornell, M.~E., \& Bender, R.\ 2009, \apjs, 182, 216 



\bibitem[Larson et al.(1980)]{larson1980} Larson, R.~B., Tinsley, B.~M., \& Caldwell, C.~N.\ 1980, \apj, 237, 692 
\bibitem[Lewis et al.(2002)]{lewis02} Lewis, I., Balogh, M., De Propris, R., et al.\ 2002, \mnras, 334, 673 
\bibitem[Lisker et al.(2007)]{lisker2007} Lisker, T., Grebel, E.~K., Binggeli, B., \& Glatt, K.\ 2007, \apj, 660, 1186 
\bibitem[Lisker et al.(2006a)]{lisker2006a} Lisker, T., Grebel, E.~K., \& Binggeli, B.\ 2006, \aj, 132, 497 
\bibitem[Lisker et al.(2006b)]{lisker2006b} Lisker, T., Glatt, K., Westera, P., \& Grebel, E.~K.\ 2006, \aj, 132, 2432 

\bibitem[MacArthur et al.(2003)]{macarthur2003} MacArthur, L.~A., Courteau, S., \& Holtzman, J.~A.\ 2003, \apj, 582, 689 
\bibitem[Mastropietro et al.(2005)]{mastropietro2005} Mastropietro, C., Moore, B., Mayer, L., et al.\ 2005, \mnras, 364, 607 
\bibitem[Meert et al.(2015)]{meert2015} Meert, A., Vikram, V., \& Bernardi, M.\ 2015, \mnras, 446, 3943 
\bibitem[M{\'e}ndez-Abreu et al.(2010)]{mendezabreu2010} M{\'e}ndez-Abreu, J., S{\'a}nchez-Janssen, R., \& Aguerri, J.~A.~L.\ 2010, \apjl, 711, L61 
\bibitem[M{\'e}ndez-Abreu et al.(2008)]{mendezabreu2008} M{\'e}ndez-Abreu, J., Aguerri, J.~A.~L., Corsini, E.~M., \& Simonneau, E.\ 2008, \aap, 478, 353 
\bibitem[Merritt(1984)]{merritt1984} Merritt, D.\ 1984, \apj, 276, 26 
\bibitem[M{\"o}llenhoff  \& Heidt(2001)]{mollenhoff2001} M{\"o}llenhoff, C., \& Heidt, J.\ 2001, \aap, 368, 16 
\bibitem[Montero-Dorta \& Prada(2009)]{monterodorta2009} Montero-Dorta, A.~D., \& Prada, F.\ 2009, \mnras, 399, 1106 
\bibitem[Moore et al.(1996)]{moore1996} Moore, B., Katz, N., Lake, G., Dressler, A., \& Oemler, A.\ 1996, \nat, 379, 613 
\bibitem[Mu{\~n}oz-Mateos et al.(2013)]{munozmateos13} Mu{\~n}oz-Mateos, J.~C., Sheth, K., Gil de Paz, A., et al.\ 2013, \apj, 771, 59 

\bibitem[Oman et al.(2013)]{oman13} Oman, K.~A., Hudson, M.~J., \& Behroozi, P.~S.\ 2013, \mnras, 431, 2307

\bibitem[Pedraz et al.(2002)]{pedraz2002} Pedraz, S., Gorgas, J., Cardiel, N., S{\'a}nchez-Bl{\'a}zquez, P., \& Guzm{\'a}n, R.\ 2002, \mnras, 332, L59 
\bibitem[Pe{\~n}arrubia et al.(2008)]{penarrubia2008} Pe{\~n}arrubia, J., Navarro, J.~F., \& McConnachie, A.~W.\ 2008, \apj, 673, 226 
\bibitem[Phillipps et al.(1998)]{phillips1998} Phillipps, S., Driver, S.~P., Couch, W.~J., \& Smith, R.~M.\ 1998, \apjl, 498, L119 
\bibitem[Popesso et al.(2006)]{popesso2006} Popesso, P., Biviano, A., B{\"o}hringer, H., \& Romaniello, M.\ 2006, \aap, 445, 29 
\bibitem[Prieto et al.(2001)]{prieto2001} Prieto, M., Aguerri, J.~A.~L., Varela, A.~M., \& Mu{\~n}oz-Tu{\~n}{\'o}n, C.\ 2001, \aap, 367, 405 




\bibitem[Quilis et al.(2000)]{quilis2000} Quilis, V., Moore, B., \& Bower, R.\ 2000, Science, 288, 1617 

\bibitem[Rodr{\'{\i}}guez Del Pino et al.(2014)]{rodriguezdelpino2014} Rodr{\'{\i}}guez Del Pino, B., Bamford, S.~P., Arag{\'o}n-Salamanca, A., et 
al.\ 2014, \mnras, 438, 1038 

\bibitem[Salo et al.(2015)]{salo15} Salo, H., Laurikainen, E., Laine, J., et al.\ 2015, \apjs, 219, 4 
\bibitem[S{\'a}nchez-Janssen \& Aguerri(2012)]{sanchezjanssen12} S{\'a}nchez-Janssen, R., \& Aguerri, J.~A.~L.\ 2012, \mnras, 424, 2614 
  \bibitem[S{\'a}nchez-Janssen et al.(2008)]{sanchezjanssen2008} S{\'a}nchez-Janssen, R., Aguerri, J.~A.~L., \& Mu{\~n}oz-Tu{\~n}{\'o}n, C.\ 2008, \apjl, 679, L77 
  \bibitem[Scannapieco \& Tissera(2003)]{scannapieco2003} Scannapieco, C., \& Tissera, P.~B.\ 2003, \mnras, 338, 880 
  \bibitem[Simard et al.(2011)]{simard2011} Simard, L., Mendel, J.~T., Patton, D.~R., Ellison, S.~L., 
\& McConnachie, A.~W.\ 2011, \apjs, 196, 11 
\bibitem[Sheth et al.(2010)]{sheth10} Sheth, K., Regan, M., Hinz, J.~L., et al.\ 2010, \pasp, 122, 1397

  \bibitem[Smith et al.(2009)]{smith2009} Smith, R.~J., Lucey, J.~R., Hudson, M.~J., et al.\ 2009, \mnras, 392, 1265 
  
\bibitem[Thuan \& Martin(1981)]{thuan81} Thuan, T.~X., \& Martin, G.~E.\ 1981, \apj, 247, 823   
\bibitem[Tissera et al.(2006)]{tissera2006} Tissera, P.~B., Smith Castelli, A.~V., \& Scannapieco, C.\ 2006, \aap, 455, 135 
\bibitem[Toloba et al.(2014)]{toloba14} Toloba, E., Guhathakurta, P., Peletier, R.~F., et al.\ 2014, \apjs, 215, 17 
  \bibitem[Toloba et al.(2009)]{toloba2009} Toloba, E., Boselli, A., Gorgas, J., et al.\ 2009, \apjl, 707, L17
  \bibitem[Trujillo et al.(2002)]{trujillo2002} Trujillo, I., Aguerri, J.~A.~L., Guti{\'e}rrez, C.~M., Caon, N., \& Cepa, J.\ 2002, \apjl, 573, L9 
\bibitem[Trujillo et al.(2002)]{trujillo2002} Trujillo, I., Asensio Ramos, A., Rubi{\~n}o-Mart{\'{\i}}n, J.~A., et al.\ 2002, \mnras, 333, 510 

 \bibitem[van Zee(2000)]{vanzee00} van Zee, L.\ 2000, \aj, 119, 2757 


 \end{thebibliography}
\end{document}